\documentclass[conference]{IEEEtran}
\IEEEoverridecommandlockouts

\usepackage{cite}
\usepackage{comment}
\usepackage{overpic}
\usepackage{amsmath,amssymb,amsfonts}
\usepackage{graphicx}
\usepackage{textcomp}
\usepackage{xcolor}
\usepackage{float}
\usepackage{amsthm}
\usepackage{graphicx}
\usepackage{epstopdf}
\usepackage{amsmath,bm}
\usepackage{amsfonts}
\usepackage{amssymb}
\usepackage{setspace}

\usepackage{color}
\usepackage{multirow}
\usepackage{multicol}
\usepackage{soul,xcolor}
\usepackage{algorithm}
\usepackage{algpseudocode}%

\theoremstyle{plain}

\newtheorem{lemma}{Lemma}

\newcommand{\vect}[1]{\mathbf{#1}}

\def\Htran{\mbox{\tiny $\mathrm{H}$}}
\def\Ttran{\mbox{\tiny $\mathrm{T}$}}
\def\CN{\mathcal{N}_{\mathbb{C}}} 
\def\imagunit{\mathsf{j}} 
\def\m{\rm}

\begin{document}

\title{Spatial Correlation Modeling and RS-LS Estimation of Near-Field Channels with Uniform Planar Arrays
\vspace{-12mm}
\thanks{The work by \"O. T. Demir was supported by 2232-B International Fellowship for Early Stage Researchers Programme funded by the Scientific and Technological Research Council of T\"urkiye.}
}

\author{\IEEEauthorblockN{ \"Ozlem Tu\u{g}fe Demir$^*$, Alva Kosasih$^{\dagger}$, and Emil Bj{\"o}rnson$^{\dagger}$}
\IEEEauthorblockA{ {$^*$Department of Electrical-Electronics Engineering, TOBB University of Economics and Technology, Ankara, T\"urkiye
		} \\
  {$^\dagger$Department of Computer Science, KTH Royal Institute of Technology, Stockholm, Sweden}}
  \IEEEauthorblockA{Email: ozlemtugfedemir@etu.edu.tr, kosasih@kth.se, emilbjo@kth.se
		\vspace{-4mm}}}
\maketitle

\begin{abstract}
Extremely large aperture arrays (ELAAs) can offer massive spatial multiplexing gains in the radiative near-field region in beyond 5G systems. While near-field channel modeling for uniform linear arrays has been extensively explored in the literature, uniform planar arrays—despite their advantageous form factor—have been somewhat neglected due to their more complex nature. Spatial correlation is crucial for non-line-of-sight channel modeling. Unlike far-field scenarios, the spatial correlation properties of near-field channels have not been thoroughly investigated. In this paper, we start from the fundamentals and develop a near-field spatial correlation model for arbitrary spatial scattering functions. Furthermore, we derive the lower-dimensional subspace where the channel vectors can exist. It is based on prior knowledge of the three-dimensional coverage region where scattering clusters exists and we derive a tractable one-dimensional integral expression. This subspace is subsequently employed in the reduced-subspace least squares (RS-LS) estimation method for near-field channels, thereby enhancing performance over the traditional least squares estimator without the need for having full spatial correlation matrix knowledge.
\end{abstract}

\begin{IEEEkeywords}
Extremely large-scale MIMO,  reduced-subspace least-square estimator, spatial correlation, near-field channels.%
\end{IEEEkeywords}

\section{Introduction}

Extremely large aperture arrays (ELAAs) emerge as a promising technology poised to meet the escalating demands for communications and support new applications in 6G and beyond wireless systems \cite{2019_Rappaport_Access,2019_Björnson_DSP}. The primary advantages offered by ELAAs include enhanced spatial multiplexing, superior interference management, and unprecedented beamforming gains \cite{9903389,hu2018beyond, demir2022channel, 10273772}. Utilizing ELAAs not only involves the integration of a larger number of antennas but also necessitates reevaluating the electromagnetic distinctions between the far-field and near-field regions, demarcated by the Fraunhofer distance. This distance increases with the square of the aperture length for a given carrier frequency, making it likely for user equipments (UEs) and/or scatterers to reside within the radiative near-field of an ELAA, even in the sub-6 GHz band \cite{2024_Kosasih_TWC}.

In conventional multiple-input multiple-output (MIMO) systems, understanding spatial correlation across channel realizations at different antennas is crucial for accurate channel modeling \cite{Sayeed2002a}. Spatial correlation cannot be ignored except in cases of isotropic scattering with isotropic antennas and when utilizing a uniform linear array (ULA) with inter-antenna spacing at integer multiples of half the wavelength. Thus, spatial correlation is typically inevitable. In far-field systems, one widely used model of spatial correlation is based on correlated Rayleigh fading under rich scattering conditions as characterized by the spatial scattering function \cite{Sayeed2002a}. Recently, several studies have extended this analysis to radiative near-field channels \cite{9763525,10437851}, though these primarily consider ULAs at the base station (BS) and focus on a specific one-ring model.

This paper builds on the fundamentals established in \cite{Sayeed2002a} to construct a spatial correlation matrix for a correlated Rayleigh fading near-field channel, considering a uniform planar array  (UPA) at the BS. We derive a spatial correlation matrix that involves triple integrals over the azimuth, elevation, and distance domains. We propose an analytically tractable method to reduce the triple integral to a single integral. This method computes a subspace where any plausible channel can be represented as a vector within this subspace since a channel realization usually has a smaller effective dimension than the number of antennas due to spatial oversampling in a UPA \cite{demir2022channel}  and limited scattering. This subspace is derived without explicit knowledge of the exact spatial correlation matrix but based on the domain where the spatial scattering function is non-zero. 
This subspace representation can be utilized in the reduced-subspace least-squares (RS-LS) estimator \cite{demir2022channel}.
We then enhance the conventional RS-LS estimator by dynamically estimating the average channel gain in each subspace dimension.
Our simulation results reveal that a better RS-LS estimate is obtained by considering both angular and distance domains when characterizing the subspace, compared to only considering angles as in far-field scenarios. The benefit becomes more pronounced as the possible distances of scatterers decrease, highlighting the importance of computing the subspace in a near-field-aware manner.

\section{System and Channel Modeling}

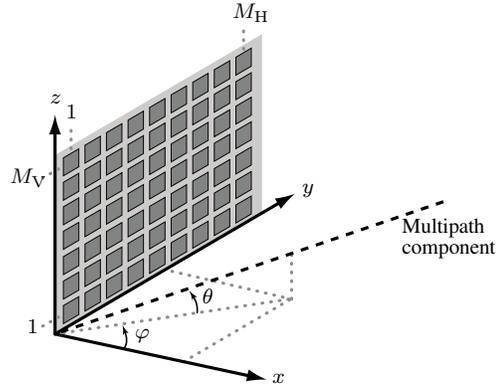
\begin{figure}[t!]
\vspace{5mm}
 \centering 
	\begin{overpic}[width=.75\columnwidth,tics=10]{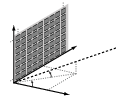}
		\put(54,1){\footnotesize $x$}
		\put(60,39){\footnotesize $y$}
		\put(9.5,57){\footnotesize $z$}
		\put(26.5,10){\footnotesize  $\varphi$}
		\put(40,17){\footnotesize  $\theta$}
		\put(5,11){\footnotesize  $1$}
		\put(1.5,41){\footnotesize  $M_{\mathrm{V}}$}
		\put(12.7,54){\footnotesize  $1$}
		\put(46,74){\footnotesize  $M_{\mathrm{H}}$}
		\put(80,31){\footnotesize  Multipath}
		\put(80,27){\footnotesize  component}
\end{overpic} 
\vspace{-4mm}
	\caption{The 3D geometry of a BS array consisting of $M_{\mathrm{H}}$ elements per row and $M_{\mathrm{V}}$ elements per column.}\vspace{-4mm}
	\label{figure_geometric_setup}  
 \vspace{-1mm}
\end{figure}

We consider a communication setup wherein a BS communicates with a UE. The BS is equipped with a large number of antennas arranged in a UPA configuration with the number of antennas denoted as $M$. This setup is illustrated in Fig.~\ref{figure_geometric_setup}. The number of antennas per row and per column of the UPA is denoted as $M_{\rm H}$ and $M_{\rm V}$, respectively, resulting in $M=M_{\rm H}M_{\rm V}$.  We consider identical and uniform spacing of $\Delta$ between adjacent antennas in both vertical and horizontal directions.  The antennas are sequentially indexed row by row, with the index parameter $m \in \{1,\dots, M\}$. Hence, the position of the $m$-th antenna relative to the origin is given by the vector $\vect{u}_m = [ 0, \, \,\, i_m \Delta,  \,\,\, j_m \Delta]^{\Ttran}$, where $i_m =\mathrm{mod}(m-1,M_{\m H})$ and $j_m =\left\lfloor(m-1)/M_{\m H}\right\rfloor$ represent  the horizontal and vertical indices, respectively. Here, $\mathrm{mod}(\cdot,\cdot)$ is the modulus operation, while $\lfloor \cdot \rfloor$ is the truncation operation.

 The Fraunhofer (Rayleigh) distance, given by $2D^2/\lambda$ with $D=\sqrt{M_{\rm H}^2+M_{\rm V}^2}\Delta$ being the aperture length, serves as the classical boundary distinguishing between the far-field and radiative near-field regions  of an array. When the UPA is extremely large, the Fraunhofer distance extends to hundreds of meters \cite{10273772}. Therefore, this paper focuses on the scenario where the scatterers are situated within the (radiative) near-field region, i.e., the Fresnel region, of the extremely large UPA. Furthermore, we consider a non-line-of-sight (NLOS) channel setting. Indeed, in this context, the distance from the UPA to each scatterer falls within the range between the Fraunhofer distance $2D^2/\lambda$ and the Fresnel distance $0.62\sqrt{D^3/\lambda}$.

 In the near field, an incident wave in the uplink exhibits a spherical wavefront. The spherical wavefront is characterized not only by the direction of the wave but also by the distance between the source and the receiver. Therefore, a path from a scatterer can be characterized using the near-field array response vector $\vect{b}\left(\varphi, \theta, r\right)$, where $r$ represents the distance from the origin (corner point of the UPA) to the respective scatterer \cite{cui2022channel}. Notice that the near-field array response vector $\vect{b}\left(\varphi, \theta, r\right)$ also depends on the azimuth and elevation angles $\varphi$ and $\theta$, which are computed with respect to the origin. The near-field array response vector characterizes the specific spherical wave and is written for the considered UPA as
\begin{equation} \label{eq:near-field-array-response-vector}
\vect{b}\left(\varphi, \theta, r\right) = \left[e^{-\imagunit\frac{2\pi}{\lambda}(r_{1}-r)}, \ldots, e^{-\imagunit\frac{2\pi}{\lambda}(r_{M}-r)} \right]^{\Ttran}, 
    \end{equation}
where $r_{m}$ denotes the distance from the BS antenna $m$ to the scatterer and it is given as 
 \vspace{-2mm}
\begin{align} \label{eq:distance-difference}
r_{m}=\bigg(&\Big(r\cos(\theta)\cos(\varphi)-0\Big)^2
+\Big(r\cos(\theta)\sin(\varphi)-i_m\Delta\Big)^2\nonumber\\
&+\Big(r\sin(\theta)-j_m\Delta\Big)^2\bigg)^{\! \frac12} \nonumber \\
&\hspace{-12mm}=r\Bigg( 1-2\Delta\frac{i_m\cos(\theta)\sin(\varphi)+j_m\sin(\theta)}{r}+\Delta^2\frac{i^2_m+j^2_m}{r^2}\Bigg)^{\! \frac12} \!.
\end{align}
We assume  that the waves arrive only from the directions in front of the array, limiting $\varphi$ to the range $[-\frac{\pi}{2},\frac{\pi}{2}]$.

To simplify \eqref{eq:distance-difference}, one can utilize the Fresnel approximation \cite{ziomek1993three}. This is done by performing a first-order Taylor approximation (i.e., $\sqrt{1+x}\approx 1+\frac{1}{2}x$, for small $x$). Utilizing this approximation, we can express $r_{m}$ in \eqref{eq:distance-difference} as  
\vspace{-2mm}
\begin{align}  \label{eq:near-field-expansion}
r_{m}&\approx  r -\Delta\Big(i_m\cos(\theta)\sin(\varphi)+j_m\sin(\theta)\Big)\nonumber\\
&\quad+\Delta^2\left(\frac{i^2_m+j^2_m}{2r}\right).
\end{align} 
When $r$ is beyond the Fraunhofer distance, the last term involving $1/r$ in \eqref{eq:near-field-expansion} can be omitted, and the near-field array response vector in \eqref{eq:near-field-array-response-vector} becomes identical to the corresponding far-field array response vector. Inserting \eqref{eq:near-field-expansion} into \eqref{eq:near-field-array-response-vector}, we obtain the approximation of the $m$th entry of $\vect{b}(\varphi,\theta,r)$ as
\begin{align}  \label{eq:proposed-approximation}
\exp\!\Bigg(\!\imagunit\frac{2\pi}{\lambda}\!\Bigg[&\!\Delta  \Big(\!i_m\cos(\theta)\sin(\varphi)\!+\!j_m\sin(\theta)\Big) \!-\! \Delta^2 \frac{i^2_m\!+\! j^2_m}{2r}\Bigg]\Bigg).
\end{align}
Building on the approximate array response vector for the scatterers outlined above, we will next analyze the near-field channel and spatial correlation models in the subsequent section, thereby rendering the analysis more tractable.

\section{Spatial Correlation in Near-Field Channels}

We assume that the UE and/or scatterers are located in the radiative near-field region of the BS array.\footnote{Many propagation environments feature a limited number of scattering clusters, and the number of clusters reduces as we move to higher frequencies.} It is known that the scatterers exist in a three-dimensional region characterized by the distance range from  $d_1$ to $d_2$, where $d_2$ is greater than $d_1$. The angular span of these clusters in the azimuth and elevation domains are from $\varphi_1$ to $\varphi_2$, and from $\theta_1$ to $\theta_2$, respectively.  The superposition of these beams is represented by the channel vector $\vect{h}\in \mathbb{C}^M$, which is given by
\begin{equation} \label{eq:channel1}
\vect{h} =   \int_{d_1}^{d_2}\int_{\theta_1}^{\theta_2}\int_{\varphi_1}^{\varphi_2} g(\varphi,\theta,d) \vect{b}(\varphi,\theta,d) \partial \varphi \partial\theta \partial d, 
\end{equation}
where the angular and distance spreading function $g(\varphi,\theta,d)$\footnote{The function is a generalization of the classical angular spreading function \cite{Sayeed2002a,demir2022channel} to the near-field channels.} determines the gain and phase-shift originating from a scatterer location $(\varphi,\theta,d)$. Similarly to the conventional far-field angular spreading function \cite{Sayeed2002a}, we model $g(\varphi,\theta,d)$ as a spatially uncorrelated circularly symmetric Gaussian stochastic process with cross-correlation
\vspace{-0.3mm}
\begin{align} \label{eq:scattering-correlation-model}
&\mathbb{E} \{ g(\varphi,\theta,d) g^*(\varphi',\theta',d') \}\nonumber\\
&= \beta f(\varphi,\theta,d) \delta (\varphi-\varphi')\delta(\theta-\theta')  \delta(d-d'),
\end{align}
where $\delta(\cdot)$ is the Dirac delta function, $\beta$ denotes the average channel gain, $f(\varphi,\theta,d)$ is the normalized \emph{spatial scattering function} (i.e., $\int_{d_1}^{d_2}\int_{\theta_1}^{\theta_2} \int_{\varphi_1}^{\varphi_2} f(\varphi,\theta,d)\partial\varphi \partial \theta \partial d=1$), which is a near-field equivalent of the far-field spatial scattering function \cite{Sayeed2002a}, and $(\varphi',\theta',d')$ represents any arbitrary combination of azimuth and elevation angles, and distance.  Therefore, the correlated Rayleigh fading channel in \eqref{eq:channel1} can be modeled as
\begin{equation} \label{eq:corr-Rayleigh}
\vect{h} \sim \CN(\vect{0},\vect{R}),
\end{equation}
which is fully characterized by the spatial correlation matrix
\begin{align} \label{eq:spatial-correlation}
&\vect{R} = \mathbb{E}\{ \vect{h} \vect{h}^{\Htran} \}= \beta  \int_{d_1}^{d_2}\int_{\theta_1}^{\theta_2}\int_{\varphi_1}^{\varphi_2} f(\varphi,\theta,d) \nonumber\\ & \cdot\vect{b}(\varphi,\theta,d)\vect{b}^{\Htran}(\varphi,\theta,d) \partial \varphi \partial\theta \partial d. 
\end{align}
Notice that each diagonal entry of $\vect{R}$ equals $\beta$, so the near-field propagation conditions do not create power variations between the antennas.

\section{Low-Dimensional Subspace Representation of Near-Field Channels}

In this section, we will analyze a general low-dimensional representation of the near-field channels. Such a representation will help us to estimate the near-field channel efficiently, as will be discussed later in Section~\ref{section:RSLS}. We begin by defining an arbitrary correlation matrix denoted as $\overline{\vect{R}}$. This matrix does not represent a specific channel but encapsulates the general array geometry and all possible scatterer locations in a given coverage region. It should be noted that $\overline{\vect{R}}$ is distinct from the exact correlation matrix $\vect{R}$ defined in  \eqref{eq:spatial-correlation}. We further define a matrix  $\overline{\vect{U}}\in \mathbb{C}^{M\times \overline{r}}$, where each column corresponds to an eigenvector of $\overline{\vect{R}}$ associated to a non-zero eigenvalue.\footnote{Throughout this paper, the rank is calculated as the effective rank, which retains a fraction $1-10^{-6}$ of the total sum of all eigenvalues. \label{footnote1}} This implies that the  matrix $\overline{\vect{R}}$ has $\overline{r}$ non-zero eigenvalues, where $\overline{r}<M$. Therefore, we can express any channel as $\vect{h}=\overline{\vect{U}}\vect{x}$ for some reduced dimension $\vect{x}\in \mathbb{C}^{\overline{r}}$. 
The following lemma is the near-field generalization of \cite[Lem.~3]{demir2022channel} and shows a way to compute $\overline{\vect{R}}$ for a given three-dimensional coverage region where all the possible scattering clusters are known to exist.

\begin{lemma} \label{lemma:span} Let $\overline{\vect{R}}$ and $\vect{R}$ be two spatial correlation matrices obtained based on the same array geometry. The spatial scattering functions corresponding to these correlation matrices are denoted by $\overline{f}(\varphi,\theta,d)$  and $f(\varphi,\theta,d)$, respectively, which are only non-zero for $\varphi\in[\varphi_1,\varphi_2]$, $\theta\in[\theta_1,\theta_2]$ and $d\in[d_1,d_2]$, enclosing all possible scatterer locations. Assume that $\overline{f}(\varphi,\theta,d)$  and $f(\varphi,\theta,d)$ are either continuous at each point on its domain or contain Dirac delta functions.

If $\overline{f}(\varphi,\theta,d)$ is non-zero for $\varphi\in[\varphi_1,\varphi_2]$, $\theta\in[\theta_1,\theta_2]$, and $d\in[d_1,d_2]$, then the subspace spanned by the columns of $\overline{\vect{R}}$ contains the subspace spanned by the columns of $\vect{R}$.
\vspace{-2mm}
\begin{proof}
The proof extends the proof of \cite[Lem.~3]{demir2022channel} to the near-field case using the spatial scattering function  $f(\varphi,\theta,d)$.
\end{proof}
\end{lemma}

According to Lemma~\ref{lemma:span}, we can find a representative $\overline{\vect{R}}$ by selecting an arbitrary spatial scattering function $\overline{f}(\varphi,\theta,d)$ that is non-zero $\forall \varphi\in[\varphi_1,\varphi_2]$, $\theta\in[\theta_1,\theta_2]$, and $d\in [d_1, d_2]$. To compute the triple integral in \eqref{eq:spatial-correlation} in a tractable and efficient manner, we can select the spatial scattering function as
\begin{align}
\overline{f}(\varphi,\theta,d)&= \underbrace{\frac{d_1d_2}{d_2-d_1}\frac{1}{\left(\sin(\theta_2)-\sin(\theta_1)\right)\left(\sin(\varphi_2)-\sin(\varphi_1)\right)}}_{=c}\nonumber\\
&\quad\cdot\frac{\cos(\varphi)\cos(\theta)}{d^2}.
\end{align}
 The constant $c$ ensures that $\int_{d_1}^{d_2}\int_{\theta_1}^{\theta_2}\int_{\varphi_1}^{\varphi_2} \overline{f}(\varphi,\theta,d)\partial \varphi\partial \theta \partial d=1$.
Substituting $\overline{f}(\varphi,\theta,d)$ into \eqref{eq:spatial-correlation}, we obtain the $(n,m)$th entry of the representative spatial correlation matrix $\overline{\vect{R}}$ as shown in \eqref{eq:R-overline} at the top of the next page.
\begin{figure*}
\begin{align}
     \left[\overline{\vect{R}}\right]_{n,m}& = c\beta \int_{d_1}^{d_2}\int_{\theta_1}^{\theta_2}\int_{\varphi_1}^{\varphi_2} \frac{\cos(\varphi)\cos(\theta)}{d^2}  e^{\imagunit \frac{2 \pi}{\lambda} \left( \Delta(i_n-i_m)\cos(\theta)\sin(\varphi)+\Delta(j_n-j_m)\sin(\theta)  - \frac{\Delta^2\left(i_n^2+j_n^2-i_m^2-j_m^2\right) }{2d} \right)} \partial \varphi \partial \theta \partial d \nonumber \\
     &=c\beta \int_{\theta_1}^{\theta_2}e^{\imagunit\frac{2\pi}{\lambda}\Delta(j_n-j_m)\sin(\theta)}\int_{\varphi_1}^{\varphi_2}e^{\imagunit\frac{2\pi}{\lambda}\Delta(i_n-i_m)\cos(\theta)\sin(\varphi)}\cos(\varphi)\cos(\theta)\left(\int_{d_1}^{d_2}\frac{1}{d^2}e^{-\imagunit \frac{ \pi}{\lambda}\frac{\Delta^2\left(i_n^2+j_n^2-i_m^2-j_m^2\right)   }{d} }\partial d \right) \partial \varphi \partial \theta \nonumber \\
     &= c\beta \int_{\theta_1}^{\theta_2}e^{\imagunit\frac{2\pi}{\lambda}\Delta(j_n-j_m)\sin(\theta)}\int_{\varphi_1}^{\varphi_2}e^{\imagunit\frac{2\pi}{\lambda}\Delta(i_n-i_m)\cos(\theta)\sin(\varphi)}\cos(\varphi)\cos(\theta) \partial \varphi \partial \theta \nonumber\\
     &\quad \cdot \underbrace{\frac{-\imagunit\lambda}{\pi\Delta^2\left(i_n^2+j_n^2-i_m^2-j_m^2\right)   }\left(e^{-\imagunit \frac{ \pi}{\lambda}\frac{\Delta^2\left(i_n^2+j_n^2-i_m^2-j_m^2\right)    }{d_2} }-e^{-\imagunit \frac{ \pi}{\lambda}\frac{\Delta^2\left(i_n^2+j_n^2-i_m^2-j_m^2\right)    }{d_1} } \right)}_{\triangleq s_{n,m}} \nonumber\\
     &\hspace{-12mm}= \begin{cases}c\beta s_{n,m}\frac{-\imagunit \lambda}{2\pi\Delta(i_n-i_m)}\int_{\theta_1}^{\theta_2}e^{\imagunit\frac{2\pi}{\lambda}\Delta(j_n-j_m)\sin(\theta)}\left(e^{\imagunit\frac{2\pi}{\lambda}\Delta(i_n-i_m)\cos(\theta)\sin(\varphi_2)}-e^{\imagunit\frac{2\pi}{\lambda}\Delta(i_n-i_m)\cos(\theta)\sin(\varphi_1)}\right)\partial \theta, & i_n\neq i_m \\
     c\beta s_{n,m}\left(\sin(\varphi_2)-\sin(\varphi_1)\right)\frac{-\imagunit\lambda}{2\pi\Delta (j_n-j_m)}\left(e^{\imagunit \frac{2\pi}{\lambda}(j_n-j_m)\sin(\theta_2)}-e^{\imagunit \frac{2\pi}{\lambda}(j_n-j_m)\sin(\theta_1)}\right), & i_n=i_m
     \end{cases}.
 \label{eq:R-overline}\end{align}
\hrulefill
\end{figure*}
The final one-dimensional integral in \eqref{eq:R-overline} can be evaluated numerically.
In conclusion, any channel vector $\vect{h}$, as characterized by the spatial correlation matrix defined in \eqref{eq:spatial-correlation}, lies in the lower-dimensional subspace,  spanned by the eigenspace of the computed matrix $\overline{\vect{R}}$, as shown in Lemma~\ref{lemma:span}. The eigenspace of the matrix $\overline{\vect{R}}$ is denoted by the semi-unitary matrix $\overline{\vect{U}}\in \mathbb{C}^{M\times \overline{r}}$, which is the eigenvector matrix corresponding to non-zero eigenvalues of $\overline{\vect{R}}$. Specifically, we express $\vect{h}$ as $\overline{\vect{U}}\vect{x}$, where $\vect{x}$ is a vector in $\mathbb{C}^{\overline{r}}$ of reduced dimension $\overline{r}<M$. 
Notably, the proposed framework above replaces the cumbersome triple integral computation with a one-dimensional integral, which highlights our main contribution to computing the correlation matrix $\overline{\vect{R}}$. In the following section, we will describe a channel estimation method using the derived correlation matrix, which does not require knowledge of the specific spatial correlation matrix.

\section{RS-LS Channel Estimation} 
\label{section:RSLS}

We first describe several channel estimation schemes, then, explain how we utilize the lower-dimensional subspace to perform the channel estimation. We consider a scenario where the UE transmits a predefined pilot sequence within each coherence block and the BS uses it to estimate the channel. This could be any of the UEs served by the BS. According to \cite[Sec.~3]{massivemimobook}, the received signal at the BS is
\begin{align} \label{eq:y}
    \vect{y} = \sqrt{\rho}\vect{h} + \vect{n},
\end{align}
where $\rho>0$ is the pilot signal-to-noise ratio (SNR) and $\vect{n}\sim \CN\left(\vect{0}, \vect{I}_M \right)$.
With full statistical knowledge, the minimum mean square error (MMSE) estimator becomes the optimal method. The MMSE estimate of $\vect{h}$ is  obtained as
\begin{align}\label{eq:MMSE-estimate}
\widehat{\vect{h}}_{\rm MMSE} = \sqrt{\rho}\vect{R}\left(\rho\vect{R}+\vect{I}_M\right)^{-1}\vect{y},
\end{align}
and requires knowledge of the matrix $\vect{R}$, defined in \eqref{eq:spatial-correlation}. 
Computing the MMSE estimate is challenging in practice due to the difficulty in acquiring the correlation matrix $\vect{R}$, which contains $M^2$ entries. This is particularly challenging when $M$ is large and/or when UE/BS transmits a small data packet.

A more practical approach is to employ the LS estimator that only requires knowledge of the pilot SNR, $\rho$. The LS estimator is $\widehat{ \vect{h}}_{\rm LS} = \frac{\vect{y}}{\sqrt{\rho}}$.
This estimator is overly conservative since we always know something about the antenna array and propagation environment, and this information should be used.

By utilizing $\overline{\vect{U}}$, defined in the previous section, we can eliminate the necessity of acquiring specific spatial scattering functions for every user location to compute the spatial correlation matrix as in the MMSE estimator. The RS-LS estimator from \cite{demir2022channel} is  \vspace{-2mm}
\begin{align} \label{eq:RS-LS-estimate}
    \widehat{\vect{h}}_{\rm RS-LS} = \frac{\overline{\vect{U}}\overline{\vect{U}}^{\Htran}\vect{y}}{\sqrt{\rho}},
\end{align}
which essentially implements the LS approach within the reduced-dimension subspace, and subsequently projects the estimation back into the original domain.  
We emphasize that the RS-LS channel estimation method can be applied to any UE, regardless of their individual correlation matrices. The basis vectors in $\overline{\vect{U}}$ span all channel components that can be generated by scatterers in the considered coverage region.

The RS-LS estimator can be enhanced using the following methodology. Assume that we are in the $L$-th coherence block for channel estimation and that the unknown spatial correlation statistics are static through the previous $L-1$ coherence blocks. Under this assumption, we can derive a sample estimator for the average power in each subspace dimension. Subsequently, an approximate MMSE estimator can be applied within each dimension. Finally, we project the estimated signal back into the original $M$-dimensional space. The corresponding estimate at the $L$-th coherence block is 
\begin{align}\label{eq:RSLS-estimate2}
\widehat{\vect{h}}_{\rm RS-LS}^{(L)} =  \overline{\vect{U}} \left( \frac{1}{\sqrt{\rho}} \vect{D}^{(L)} \right) \overline{\vect{U}}^{\Htran}\vect{y} 
\end{align}
where $\vect{D}^{(L)} = (\vect{\Lambda}^{(L)}-\vect{I}_{\overline{r}})   (\vect{\Lambda}^{(L)})^{-1}$ is a diagonal matrix, where $\vect{\Lambda}^{(L)}$ includes the sample estimator for the average power in each plausible channel dimension, which can be obtained by averaging the squared magnitude of the channel components $\overline{\vect{U}}^{\Htran}\vect{y}$ across $L-1$ previous received pilot signals.

\section{Numerical Results}
In this section, we analyze the eigenvalue distribution of the spatial correlation matrix for both near-field (NF) and far-field (FF) channels. We then compare the performance of the proposed RS-LS channel estimators against traditional alternatives, including MMSE and LS estimators. Two variants of the RS-LS channel estimators are depicted in the figures: RS-LS-NF, which computes the NF subspace using \eqref{eq:R-overline}, and RS-LS-FF, which computes the subspace under the FF assumption, i.e., by evaluating a similar integral as in  \eqref{eq:R-overline} but ignoring the distance variable.

We consider a UPA with $M_{\rm H}=32$ antennas per row and $M_{\rm V}=32$ antennas per column. The antenna spacing is $\Delta=0.5\lambda$, where $\lambda=0.1$\,m corresponds to a 3\,GHz carrier frequency. The aperture length is $D=\sqrt{M_{\rm H}^2+M_{\rm V}^2}\Delta \approx 2.26$\,m, leading to a Fraunhofer distance of 102.4\,m, computed as $2D^2/\lambda$. The three-dimensional region in Lemma~\ref{lemma:span} with potential scatterer locations is set as  $\varphi_1=-\pi/6$, $\varphi_2=\pi/6$, $\theta_1=-\pi/9$, and $\theta_2=0$. In the true channels, there are ten scattering cluster locations inside this region with random channel gains, angles, and distances. Hence, the corresponding spatial scattering function is composed of ten Dirac functions.

\begin{figure}[t!]
	\centering 
	\begin{overpic}[width=.825\columnwidth,tics=10]{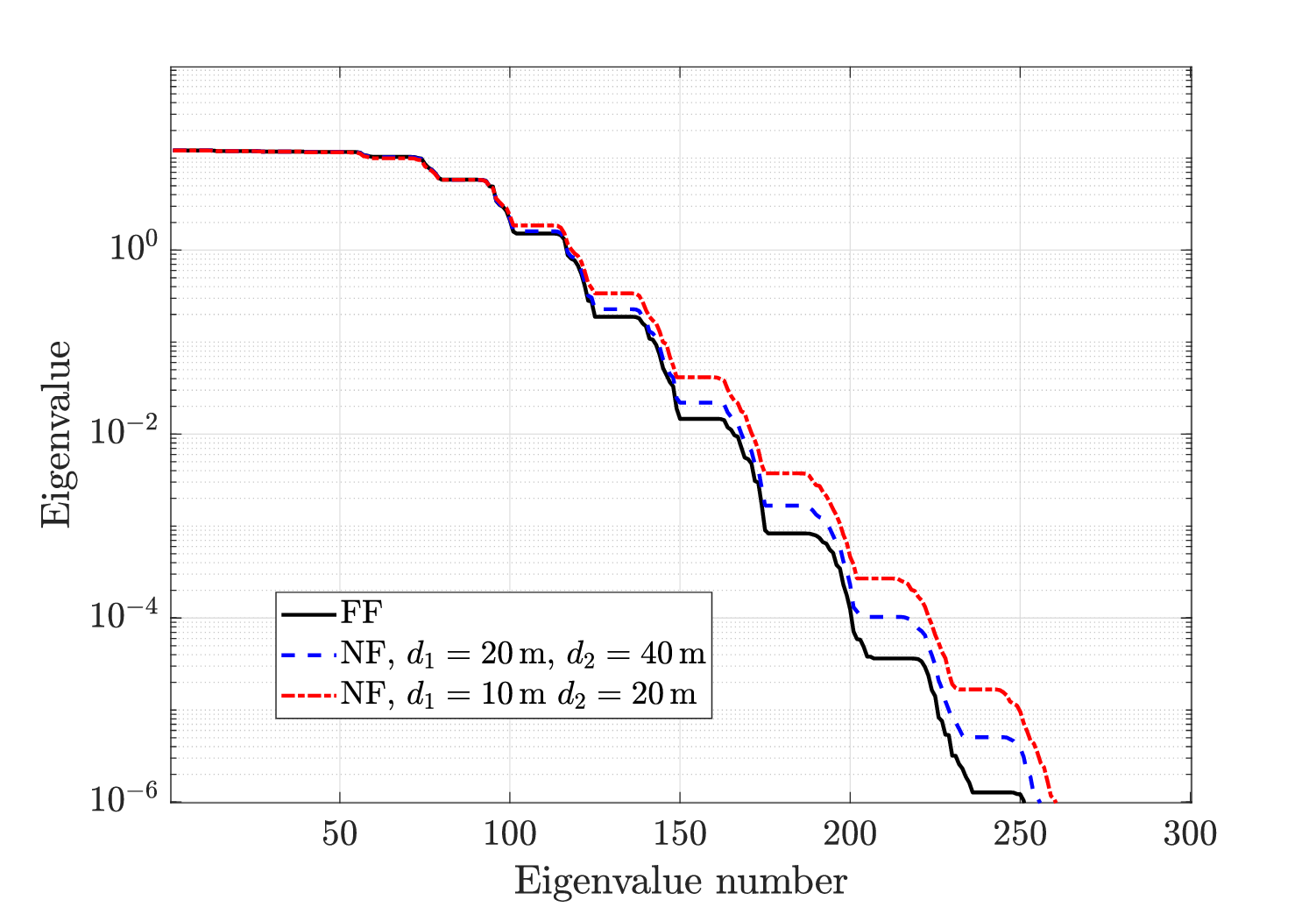}
 \end{overpic}
 \vspace{-4mm}
 \caption{The sorted eigenvalues of the spatial correlation matrices corresponding to NF and FF channels.}\vspace{-4.7mm}
	\label{simfig1}  
\end{figure}

In Fig.~\ref{simfig1}, we plot the sorted eigenvalues for the FF and NF spatial correlation matrices in two different scenarios: i) $d_1=20$\,m and $d_2=40$\,m, and ii) $d_1=10$\,m and $d_2=20$\,m. The latter scenario, with scatterers closer to the BS, exhibits a noticeable leftward shift in the NF plots, highlighting a significant energy spread effect of near-field propagation paths in the angular domain, as discussed in \cite{cui2022channel}. Consequently, the effective rank of the NF spatial correlation matrices is higher and increases as the distances decrease, indicating a mismatch between the subspaces utilized in RS-LS-NF and RS-LS-FF.

\begin{figure}[t!]
	\centering 
	\begin{overpic}[width=.825\columnwidth,tics=10]{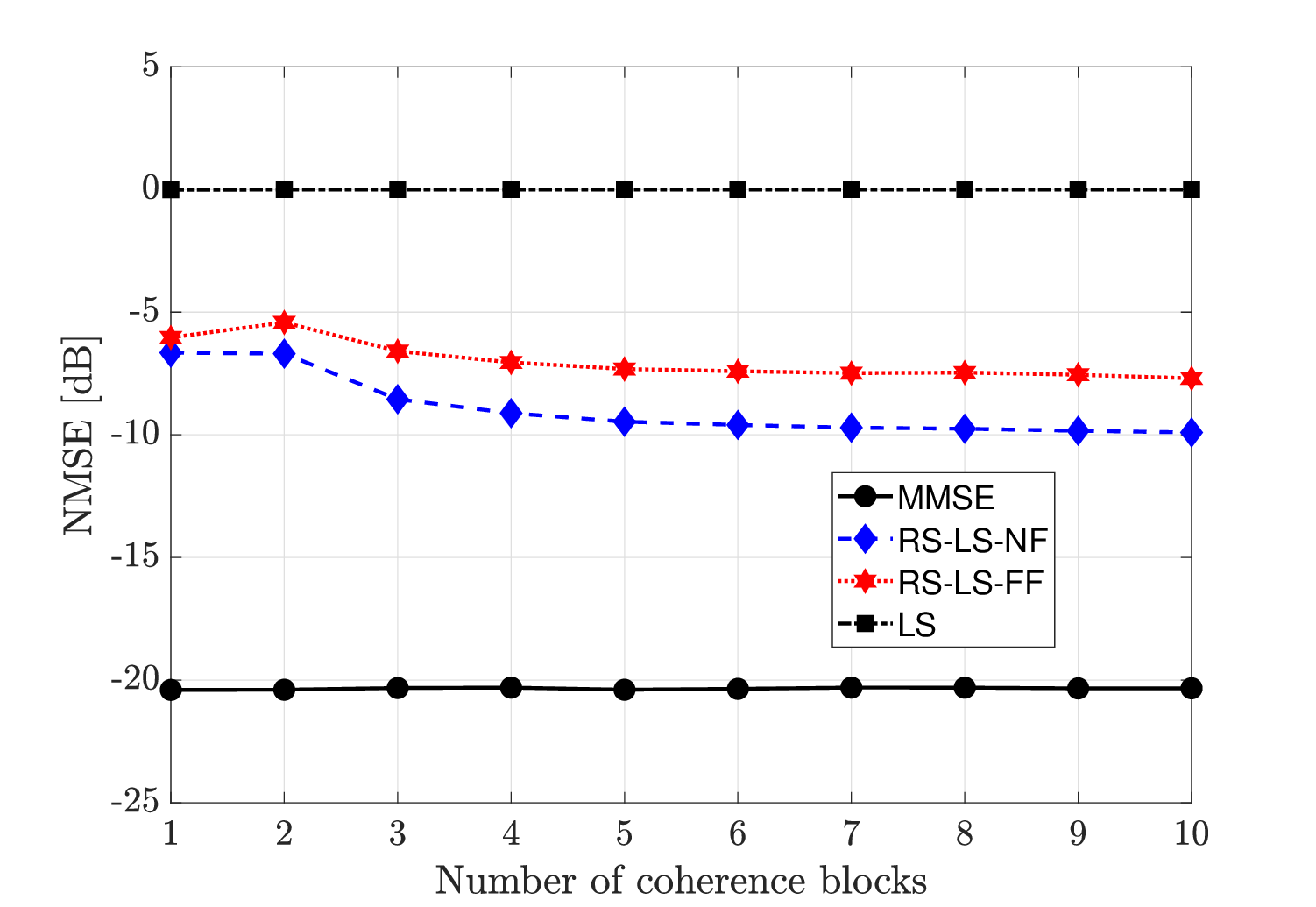}
 \end{overpic}
 \vspace{-4mm}
 \caption{The comparison of different channel estimators for SNR of $0$\,dB.}\vspace{-4mm}
	\label{simfig2}  
\end{figure}

\begin{figure}[t!]
	\centering 
	\begin{overpic}[width=.825\columnwidth,tics=10]{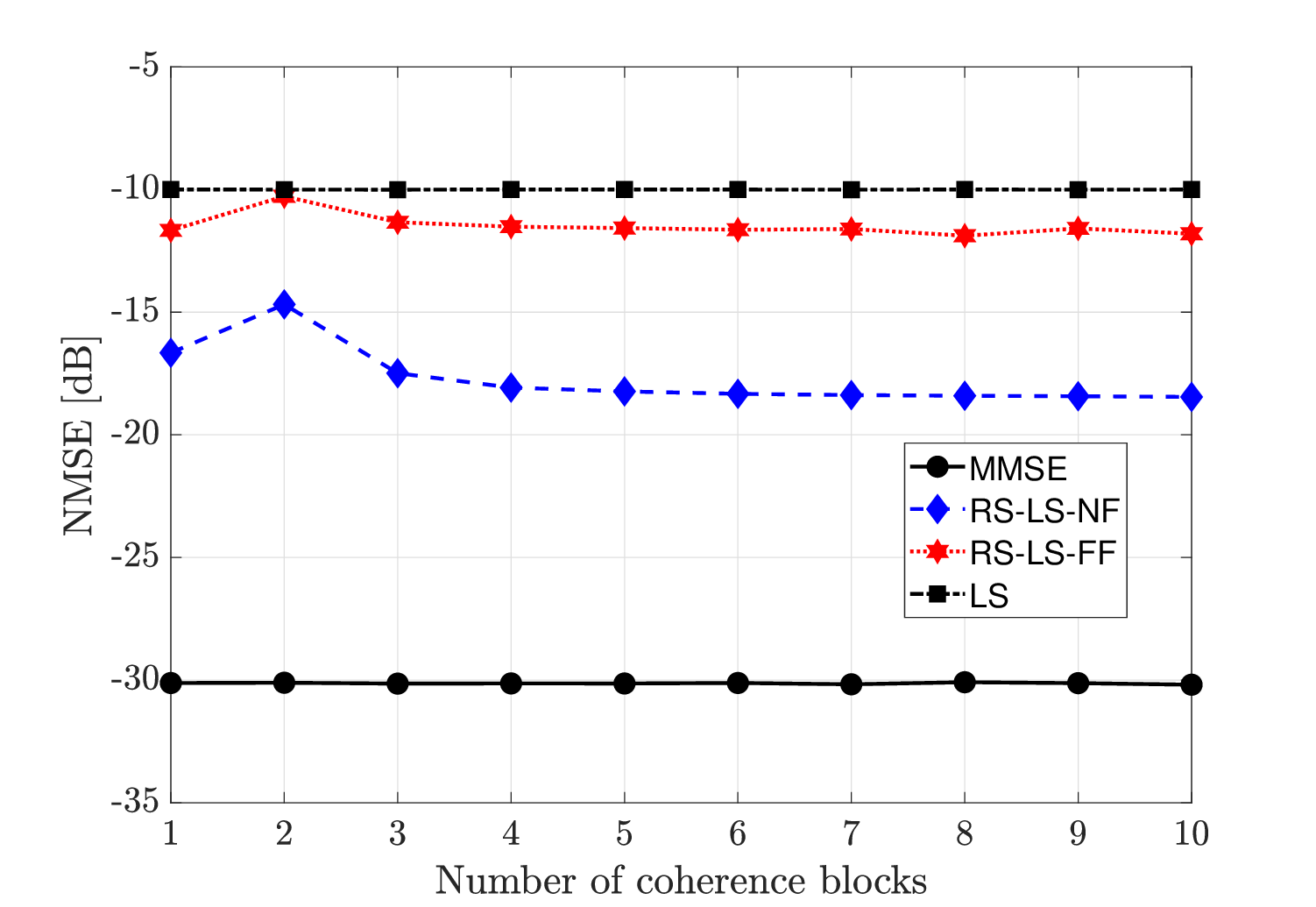}
 \end{overpic}
 \vspace{-4mm}
 \caption{The comparison of different channel estimators for SNR of $10$\,dB.}\vspace{-4.2mm}
	\label{simfig3}  
\end{figure}

We further evaluate the channel estimation performance in terms of the normalized mean-squared error (NMSE) with respect to the number of coherence blocks the estimator is used. We consider the second scenario where $d_1=10$\,m and $d_2=20$\,m. In Fig. \ref{simfig2}, the SNR is set to $0$\,dB, while in Fig. \ref{simfig3}, it is increased to $10$\,dB. As the coherence block number $L$ changes, the MMSE and LS estimators are flat since the same statistical information is utilized in the MMSE estimator. On the other hand, the proposed RS-LS estimator in \eqref{eq:RSLS-estimate2} is employed by exploiting the estimated channel gains in each dimension as the number of coherence blocks increases. Although the MMSE estimator consistently surpasses other methods, it cannot be used in practice when using ELAA since it is hard to acquire the $M^2$ entries of the spatial correlation matrix. 
A more practical benchmark is the LS estimator and it is outperformed by the proposed RS-LS-NF estimator, which exploits the spatial correlation imposed by the array geometry and coverage region without explicit knowledge of the user-specific spatial correlation matrix. Nevertheless, to ensure accuracy, it is crucial that the computation of this subspace is executed considering near-field effects to prevent underestimating the subspace dimensions---an issue evident in RS-LS-FF. This leads to notably inferior performance due to its failure to account for relevant channel dimensions, especially at higher SNRs. The proposed RS-LS-NF estimator, utilizing an eigenspace of $\overline{\vect{R}}$ as calculated in \eqref{eq:R-overline}, offers an improvement.
The performance of the RS-LS estimators improves by several dB when $L$ increases and $L=5$ is sufficient to exploit this gain.
However, since the correlation between different channel dimensions is not exploited, there remains a performance gap relative to the MMSE estimator. Nevertheless, the proposed RS-LS-NF estimator does not require specific knowledge of the spatial correlation matrix, making it easy to implement.

\section{Conclusions}

The spatial correlation in wireless channels is characterized by the array geometry and propagation environment. This becomes particularly evident when using an ELAA because radiative near-field propagation changes the channel properties.
While it is challenging to estimate large user-specific spatial correlation matrices in practice, we can characterize a subspace where all plausible channels reside. In this paper, we have derived such a subspace for a given UPA configuration and a three-dimensional coverage region containing the angles and distances of all scattering clusters. We have demonstrated how to exploit the low-dimensional subspace for RS-LS channel estimation to obtain substantially better estimation performance than with the LS estimator. An enhanced dynamic RS-LS was further proposed. We have highlighted the importance of capturing near-field properties when computing the subspace.

\vspace{-1.5mm}

\bibliographystyle{IEEEtran}
\bibliography{IEEEabrv,refs}

\end{document}